\documentclass[letterpaper,twocolumn]{jpsj3}
\usepackage{graphicx}
\usepackage{color}
\newcommand{\ltsim}{\protect\raisebox{-0.5ex}{$\:\stackrel{\textstyle <}{\sim}\:$}}

\title{Singular-Value-Decomposition Analysis of Associative Memory in a Neural Network}

\author{Tatsuya Kumamoto$^1$, Mao Suzuki$^2$, and Hiroaki Matsueda$^2$\thanks{matsueda@sendai-nct.ac.jp}}
\inst{$^1$Department of Applied Physics, Tohoku University, Sendai 980-8577, Japan \\
$^2$Sendai National College of Technology, Sendai 989-3128, Japan} 

\recdate{\today}

\abst{
We evaluate performance of associative memory in a neural network by based on the singular value decomposition (SVD) of image data stored in the network. We consider the situation in which the original image and its highly coarse-grained one by SVD are stored in the network and the intermediate one is taken as an input. We find that the performance is characterized by the snapshot-entropy scaling inherent in the SVD: the network retrieves the original image when the entropy of the input image is larger than the critical value determined from the scaling. The result indicates efficiency of the SVD as a criterion of the performance and also indicates universality of the scaling for realistic problems beyond theoretical physics.
}

\begin{document}

\maketitle

\section{Introduction}

Singular value decomposition (SVD) is an efficient way of principle component analysis. When we apply SVD to image processing, it can detect how many length-scale structures are contained in the image. Therein, the larger (smaller) components are precisely characterized by the larger (smaller) singular values and the corresponding singular vectors. Thus, the truncation by $\chi$ larger singular values leads to coarse graining. For realistic images, we know phenomenologically that the Shannon entropy defined by the singular values, called snapshot entropy, obeys logarithmic scaling with $\chi$ for relatively small $\chi$ region and then saturates toward the maximum $\chi$~\cite{Matsueda}. This means that the image is still highly coarse-grained or damaged for $\log\chi$ scaling region while the original image is almost recovered for a $\chi$ region where the entropy saturates.

In this paper, we would like to mention that this entropy scaling can measure the performance of associative memory in a neural network. Let us consider the situation in which the network memorizes two images: one is the original image, and the other is highly truncated one by SVD. We input the intermediate one, and then ask which one of memorized images is retrieved from the intermediate one. We expect that if the entropy of the intermediate image obeys the logarithmic scaling the network would retrieve the damaged one. On the other hand, the network would retrieve the original image, if the intermediate one has large enough entropy indicating the presence of fine structures of the original image. Thus, the entropy scaling behaves as a distance between the input and one of the memorized data.

There are two focus points that motivate the present neural-network study. One is to examine the functionality of the SVD in a variety of problems. This is a very general motivation associated with the SVD, not specified with the neural networks. The SVD plays central roles on recent development of tensor network variational theory for quantum many-body systems, in which the scaling feature of the SVD spectrum properly represents the amount of quantum entanglement~\cite{White,Verstraete,Vidal,Holzhey,Cardy,Andersson,Tagliacozzo,Pollmann}. This feature is deeply intertwined with renormalization group (RG) and holography concepts~\cite{Vidal,Maldacena,Takayanagi,Swingle}, and now it is very important to understand their relationship with the SVD. For this purpose, one of the present authors (HM) has been trying the SVD analysis of the snapshots at criticality of the classical spin models~\cite{Matsueda1,Matsueda2,Matsueda3,Matsueda4,Matsueda5,Matsueda6,Imura}. As will be later discussed, the neural network for the image processing is also a nice alternative approach to provide us with important insights for the scaling analysis of the SVD spectrum of images.

The other is about how to naturally evaluate network performance by the absolute measure originating in the RG concept. In comparison with the first motivation, this is rather neural-network-oriented one. Generally speaking, it is not definite to find the absolute measure for the performance of a particular neural network. Usually, it is only possible to just compare one with another network for the same problem in order to identify their superiority or inferiority. That is because each network does not have its own evaluation function, and the function is always introduced from outside of the network theory. Then, the universality of the SVD becomes very efficient to overcome such ambiguity. Since the SVD decomposes an image into a set of substructures with different length scales, the combination of the SVD with the neural network enables us to represents what kinds of length-scale data are temporary processed in the network. The crucial role of the decomposition on the performance evaluation is to introduce the extra dimension that clearly represents internal structures of the image.

The second motivation is also closely related to the rapid development of artificial intelligence based on the deep learning or the hierarchical neural networks. The presence of the extra dimension by the SVD is the emergence of the RG flow parameter in the sense of the holography, and this feature is nothing but the main function of the deep learning. Therefore, we believe that the present SVD analysis also provides us with efficient information on the advanced neural network techniques.

\section{Method}

\subsection{Neural Network Model}

In this paper, we numerically treat the most elementary Boltzmann machine. The mean-field behavior has been investigated for a long time in the context of the spin glass and also there are recent developments~\cite{Hopfield,Amit,Petersen,Tanaka,Shamir,Huang,Mezard}.

We start with the two-dimensional classical Ising-spin model as a model of recurrent neural network. The Hamiltonian is defined by
\begin{eqnarray}
H=-\sum_{i\ne j}J_{ij}\sigma_{i}\sigma_{j},
\end{eqnarray}
where $\sigma_{i}$ denotes the Ising spin or the neuron state at site $i$ and takes $\pm 1$ (up and down for the Ising spin or stationary and ignition states for the neuron). These values correspond to color degrees of freedom (black and white) in the image processing. In the next subsection, we will discuss how to extend this model to the grayscale image. The interaction between two neurons $\sigma_{i}$ and $\sigma_{j}$, $J_{ij}$, is determined by the Hebb rule:
\begin{eqnarray}
J_{ij}=\sum_{p=1}^{2}\xi_{i}^{p}\xi_{j}^{p},
\end{eqnarray}
where $\xi_{i}^{p}$ represents the color at site $i$ of the $p$-th image stored in the network. We store the original and highly coarse-grained images in the network, and they are represented as $\xi_{i}^{p=2}$ and $\xi_{i}^{p=1}$, respectively. We are going to initially input a moderately coarse-grained image $\eta$ that is intermediate between $\xi_{i}^{2}$ and $\xi_{i}^{1}$ in the sense of SVD, and then consider which one of stored two images is retrieved from the input. It is clear that the total energy tends to decrease when all of the spins take the same directions with one of $\xi_{i}^{p}$ since
\begin{eqnarray}
H=-\sum_{i\ne j}\sum_{p}\xi_{i}^{p}\xi_{j}^{p}\sigma_{i}\sigma_{j} = -\sum_{p}\left(\sum_{i}\xi_{i}^{p}\sigma_{i}\right)^{2} + {\rm const.}.
\end{eqnarray}
By this setup, the network can retrieve one of two stored images from the input data. Updating spin configuration is made by the Monte Carlo calculation at very low temperature.

\subsection{Representation of Grayscale Images by Binary Neurons}

To extend the abovementioned approach to the grayscale image with $256=2^{8}$ gradation, we have two methods. One is to define artificial $8$ neurons per pixel with a proper weight function. The second one is to use the $q$-states Potts model instead of the Ising model ($q=2$). We have confirmed that both approaches show essentially the same results for the evaluation of the network ability except for some minor difference originating in the first-order phase transition for the uniform Potts model with $q\ge 5$. The Potts-model approach is straightforward, but it has high cost on numerical simulation. Thus, we focus on the numerical results obtained by the former approach. For example, let us consider a situation in which we have $177$ as the gradation on a particular ($i$-th) pixel. The binary representation of this value is $10110001$, and each binary digit is regarded as a value of each of the artificial $8$ neurons at this pixel $i$, $\sigma_{i,\mu}$ ($\mu=1,2,...,8$. We replace $0$ into $-1$ in the calculation). For instance, if $6$ neurons of the input data $\sigma_{i,\mu}$ are equal to those of one of restored images $\xi_{i,\mu}^{p}$ (we add the index $\mu$ to $\sigma_{i}$ and $\xi_{i}^{p}$),  their product $\xi_{i,\mu}^{p}\sigma_{i,\mu}(=4)$ just represents how many binary digits match with each other. Unfortunately, we cannot distinguish different grayscales with each other. One effective method to resolve this problem is to introduce the weight function $w^{\mu}$ ($\mu=1,2,...,8$) that defines the order of importance of the $8$ neurons. It is natural that the higher digit is much more important to determine the grayscale data. Therefore, we define the weight as $128,64,32,16,8,4,2,1$ from left to right neurons of $10110001$. Namely, we assume
\begin{eqnarray}
w^{\mu}=2^{8-\mu}. \label{weighting}
\end{eqnarray}
The reliability of this assumption has been confirmed by calculating the free energy curve as a function of Hamming distance of any image from $\xi^{p=1}$. The Hamiltonian is finally represented as
\begin{eqnarray}
H &=& -\sum_{i,j}J_{ij}^{\mu\nu}\sigma_{i,\mu}\sigma_{j,\nu} \nonumber \\
&=& -\sum_{p=1}^{2}\left(\sum_{i}\sum_{\mu=1}^{8}w^{\mu}\xi_{i,\mu}^{p}\sigma_{i,\mu}\right)^{2} + {\rm const.},
\end{eqnarray}
where the coupling constant $J_{ij}^{\mu\nu}$ is defined by
\begin{eqnarray}
J_{ij}^{\mu\nu}=\sum_{p=1}^{2}\xi_{i,\mu}^{p}\xi_{j,\nu}^{p}w^{\mu}w^{\nu}.
\end{eqnarray}
By the abovementioned representation, we can apply the Monte Carlo calculation to each of $\sigma_{i,\mu}$ like the standard Boltzmann machine learning.

\subsection{Singular Value Decomposition and Snapshot Entropy}

Next we define our sample image and introduce coarse-grained images by SVD. The size of the original image is $L\times L$ (later we take $L=128$). We take the original image as $\psi(x,y)$ with $i=(x,y)$. Each two-dimensional lattice site has a value from $0$ to $255$ corresponding to grayscale data of the image. This is decomposed into SVD as
\begin{eqnarray}
\xi_{i=(x,y)}^{p=2}=\psi(x,y)=\sum_{l=1}^{L}\psi^{(l)}(x,y),
\end{eqnarray}
where
\begin{eqnarray}
\psi^{(l)}(x,y)=U_{l}(x)\sqrt{\Lambda_{l}}V_{l}(y)
\end{eqnarray}
with the singular value $\sqrt{\Lambda_{l}}$ and column unitary matrices $U_{l}(x)$ and $V_{l}(y)$. The coarse-grained image with $m$ states kept is then defined by the partial sum of $\psi^{(l)}(x,y)$ as
\begin{eqnarray}
\xi_{i=(x,y)}^{p=1}=\psi_{m}(x,y)=\sum_{l=1}^{m\ll L}\psi^{(l)}(x,y).
\end{eqnarray}
The parameter $m$ will be later determined, depending on our target images. We store both $\xi^{2}$ and $\xi^{1}$ in the network, and introduce the input $\eta$ defined by
\begin{eqnarray}
\eta(x,y)=\psi_{\chi}(x,y)=\sum_{l=1}^{\chi}\psi^{(l)}(x,y),
\end{eqnarray}
with the condition $1\ltsim m <\chi < L$.

To evaluate the resolution of $\xi^{p}(x,y)$ and $\eta(x,y)$, we normalize the SVD spectrum of the original image as
\begin{eqnarray}
\lambda_{l}=\frac{\Lambda_{l}}{\sum_{l}\Lambda_{l}},
\end{eqnarray}
and introduce the coarse-grained snapshot entropy as
\begin{eqnarray}
S_{\chi}=-\sum_{l=1}^{\chi}\lambda_{l}\ln\lambda_{l}.
\end{eqnarray}
Note that $S_{L}>S_{\chi}>S_{m}$.

\subsection{Our Strategy}

Based on the fundamentals of SVD, it would be better to make our strategy clearer by visualizing the function of neural network coupled with SVD.

\begin{figure}[htbp]
\begin{center}
\includegraphics[width=8cm]{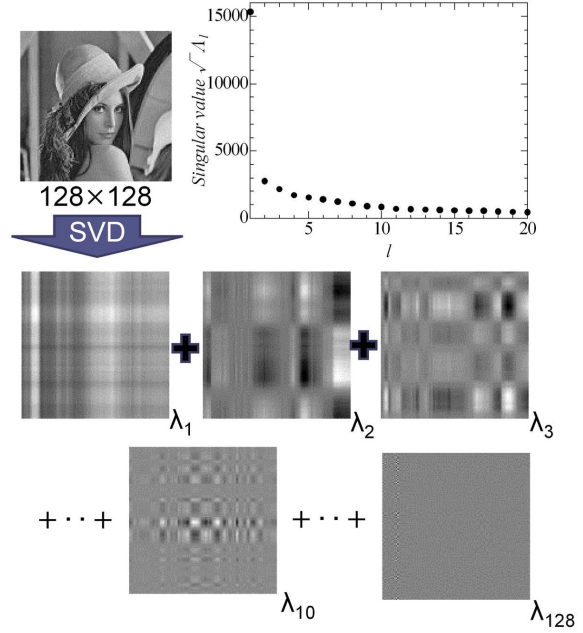}
\end{center}
\caption{SVD spectrum of LENNA image and $\psi^{l}(x,y)$ ($l=1,2,3,...,10,...,128$).}
\label{fig01}
\end{figure}

At first, we present the singular value spectrum $\sqrt{\Lambda_{l}}$ of LENNA image as a function of the SVD index $l$ in Fig.~\ref{fig01}. We see that the largest singular value $\sqrt{\Lambda_{1}}$ is exceptionally large and the spectrum is decreasing with $l$. Then, roughly $20$ singular states are important to almost recover the original LENNA image. The spectrum is related to length-scale decomposition of the original image. Actually as shown in Fig.~\ref{fig01}, a set of $\psi^{(l)}(x,y)$ shows various scale data like the wavelet transform. This feature is quite common among a large class of realistic images. The present method might have some connection with the wavelet neural network.

For small-$\chi$ region of various images, we will find the following universal scaling relation
\begin{eqnarray}
S_{\chi}\propto \log\chi, \label{scaling}
\end{eqnarray}
(see Figs.~\ref{fig03} and \ref{fig04}). It is easy to understand that this logarithmic scaling originates in algebraic decay of the SVD spectrum. According the previous works, we expect that this scaling represents complexity of the corresponding image~\cite{Matsueda,Matsueda1,Matsueda2,Matsueda3,Matsueda4,Matsueda5,Matsueda6,Imura}. As shown in Fig.~\ref{fig01}, the data space is exponentially devided into smaller pieces with increasing the SVD index $l$ to recover the original image. By taking the new coordinate $N$ by defining $\chi=a^{N}$ with a constant $a$, the snapshot entropy is proportional to $N$. This means how many length scales are contained in the image, and thus the entropy represents the complexity of the image. The entropy value deviates from Eq.~(\ref{scaling}), when $\psi_{\chi}(x,y)$ has almost recovered the original image (see Figs.~\ref{fig03} and \ref{fig04}). Therefore, this scaling determines the quality of the image $\psi_{\chi}(x,y)$ in terms of RG-like concept.

\begin{figure}[htbp]
\begin{center}
\includegraphics[width=8.5cm]{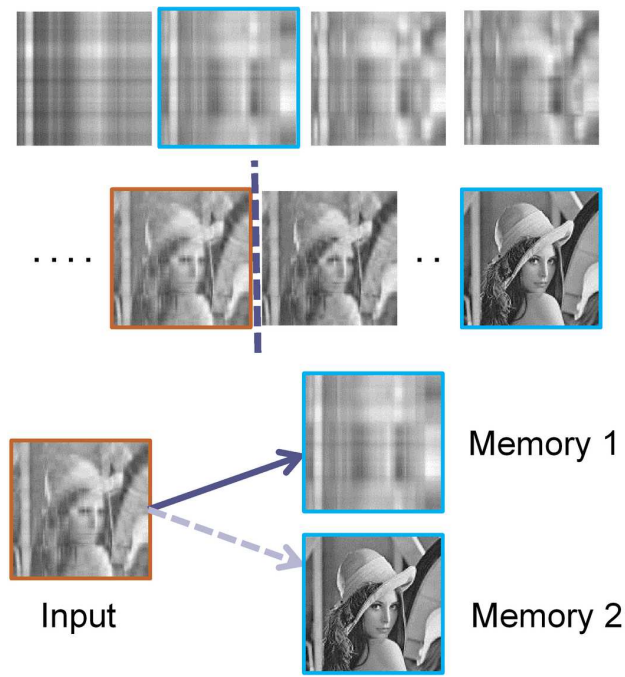}
\end{center}
\caption{Storage of two images into the network. The input data are intermediate between them, and one of which is finally remembered after Monte Carlo simulation.}
\label{fig02}
\end{figure}

Next, we show $\psi_{\chi}(x,y)$ with various $\chi$ values. In Fig.~\ref{fig02}, we align these data from left to right and top to bottom. For instance, the left upper panel represents $\psi_{1}(x,y)$. Now, our network memorizes two data: $\xi_{i=(x,y)}^{p=1}$ (Memory 1) and $\xi_{i=(x,y)}^{p=2}$ (Memory 2). Then, we select the input so that the input is located on the intermediate position between them. When the input is very similar to the original image (located at the right hand side of the dashed vertical line), the network would retrieve $\xi_{i=(x,y)}^{p=2}$. The problem is whether this consideration is well chaptured by the entropy scaling in Eq.~(\ref{scaling}).

\section{Numerical Results}

\subsection{LENNA}

\begin{figure}[htbp]
\begin{center}
\includegraphics[width=7.5cm]{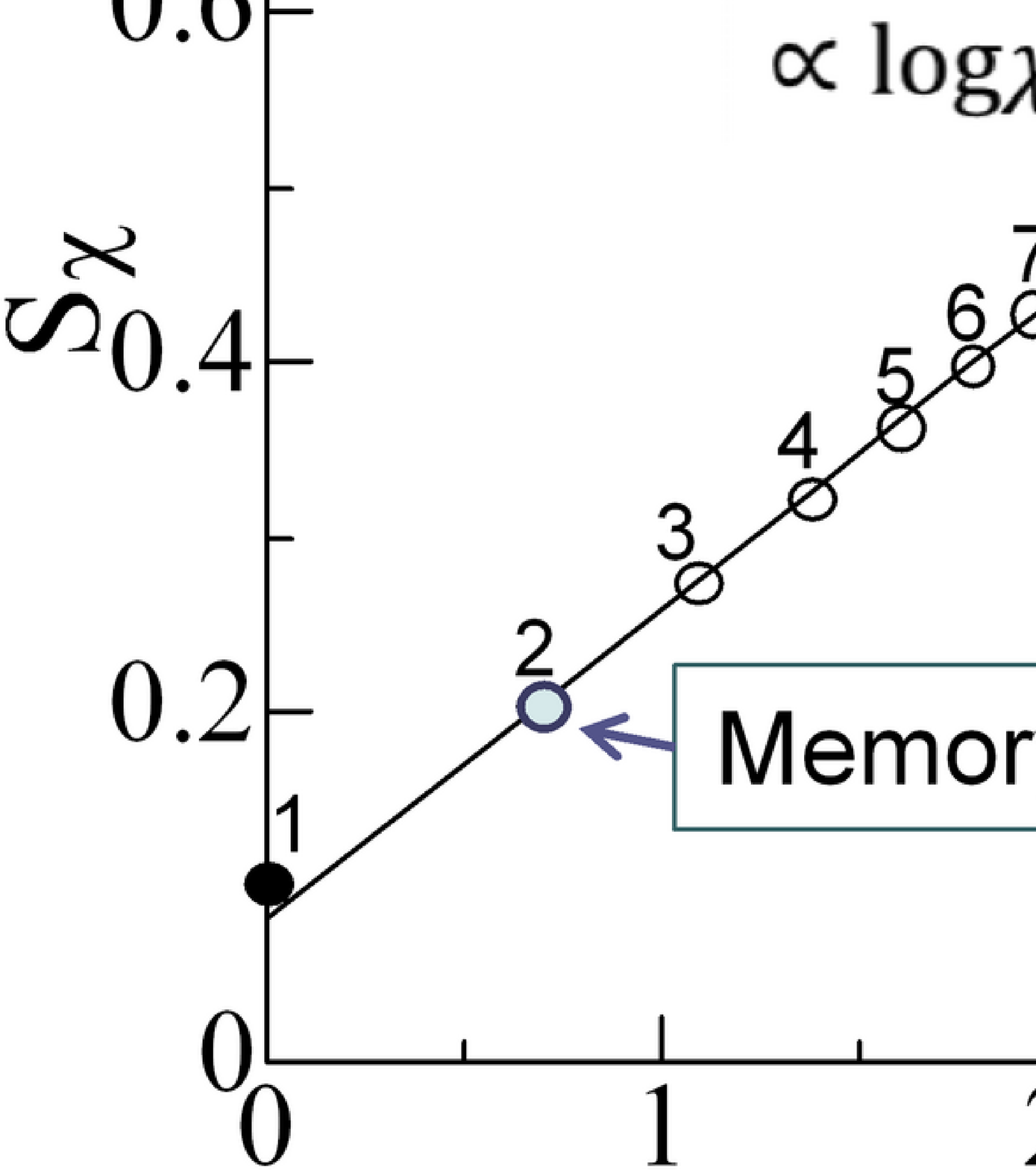}
\includegraphics[width=7.5cm]{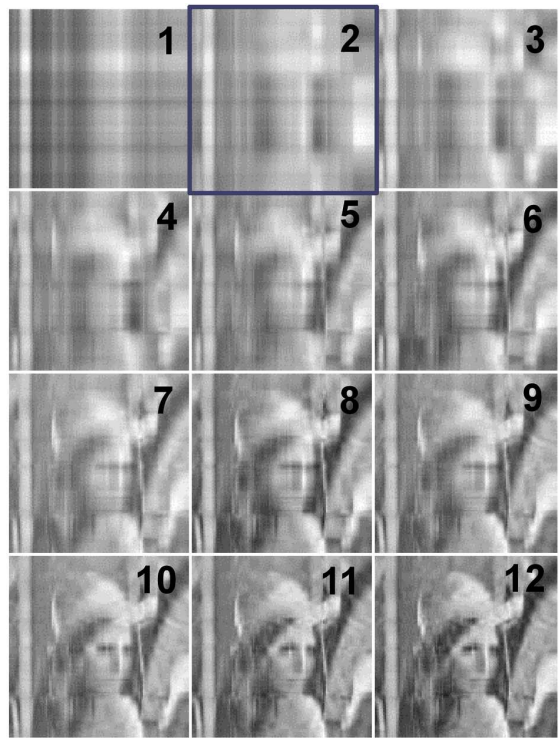}
\end{center}
\caption{(Upper panel) Snapshot entropy as a function of $\log\chi$. (Lower panel) $\psi_{\chi}(x,y)$ for $\chi=1,2,...,12$.}
\label{fig03}
\end{figure}

Let us look at numerical data for the LENNA image and consider how the network retrieves one of the stored images depending on the input. The upper panel of Fig.~\ref{fig03}(a) is the coarse-grained entropy scaling as a function of $\log\chi$, and a set of lower panels represent $\psi_{\chi}(x,y)$ for $\chi=1,2,...,12$. We assume that $m=2$ and $\xi_{i=(x,y)}^{p=1}=\psi_{m=2}(x,y)$, since $m=1$ data are almost featureless. We find that $S_{\chi}$ is proportional to $\log\chi$ for $\chi\le 10$ and tends to saturate for larger $\chi$ values. For the $\log\chi$ region, each image is highly coarse-grained, and they are in some sense similar to $\xi_{i=(x,y)}^{p=1}$. By based on this entropy scaling, we consider the performance of the network.

Here, we have plotted two types of data marked differently. The data with $\chi=3,4,5,6,7,9,10$ are marked by open circles, while the other data are marked by filled circles. This means that the original LENNA image is recovered by the network when the filled-circle data are input. Thus, the entropy scaling clearly characterizes the ability of the network. The deviation of the entropy value from the scaling line determines the critical entropy value. One exceptional data point appears for $\chi=8$. When we look at $\psi_{8}(x,y)$, the color contrast becomes slightly sharper than that for $\psi_{7}(x,y)$ and $\psi_{9}(x,y)$. This might be due to ambiguity of parametrization of grayscale by multiple neurons at each site.

It is noted that the transition between filled- and open-circle states is very strict in the present case. This is because we use the exponential weighting factor in Eq.~(\ref{weighting}). This strong weighting determines the unique critical entropy value. On the other hand, in the Potts-model approach, the probability of retrieval of the original image gradually increases with $\chi$ near the transition region. Thus, the ability of retrieval depends on the network design in general, although the snapshot entropy is still a good measure for the performance evaluation.

\subsection{Schr\"{o}dinger Cat}

\begin{figure}[htbp]
\begin{center}
\includegraphics[width=7.5cm]{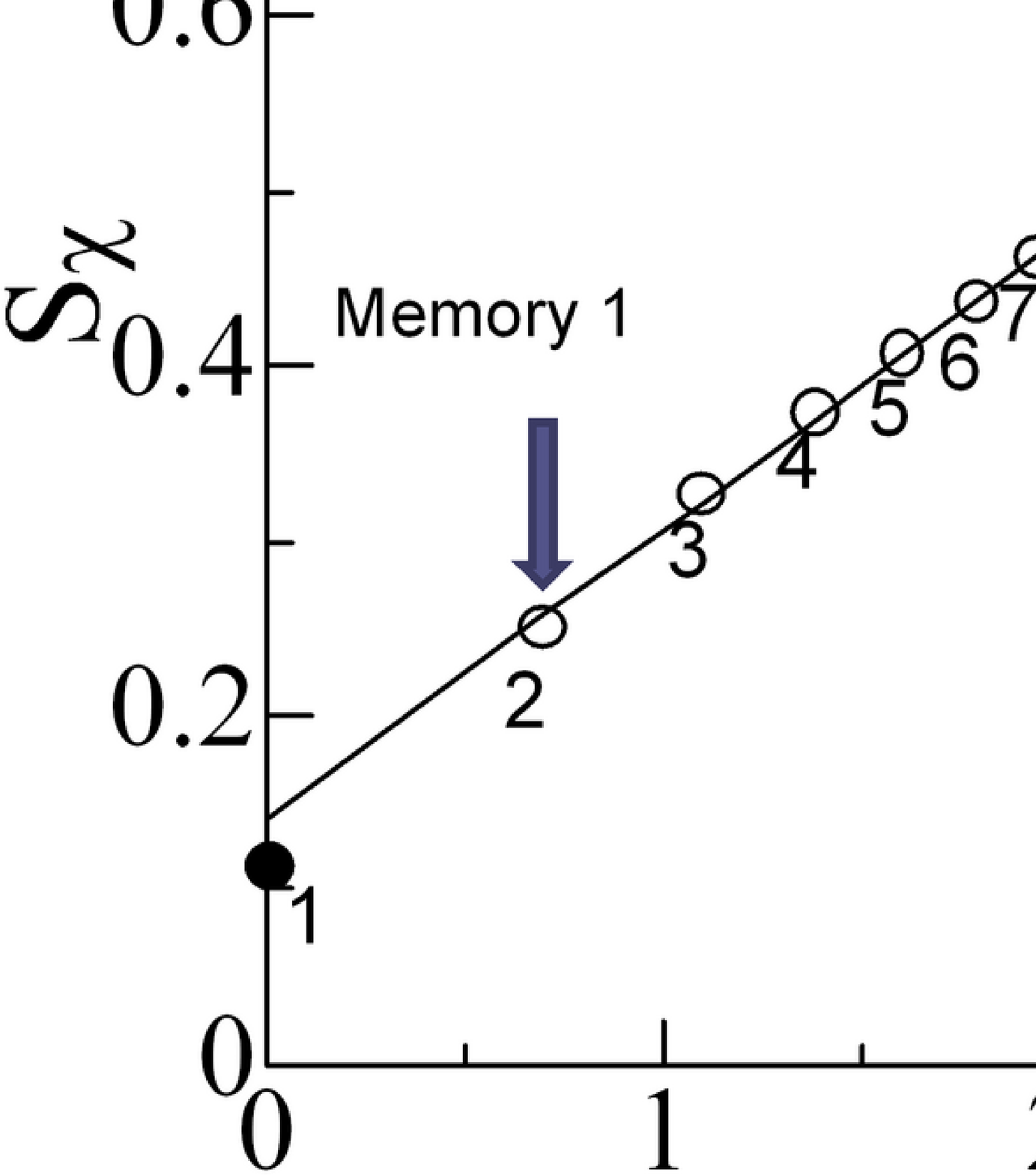}
\includegraphics[width=7.5cm]{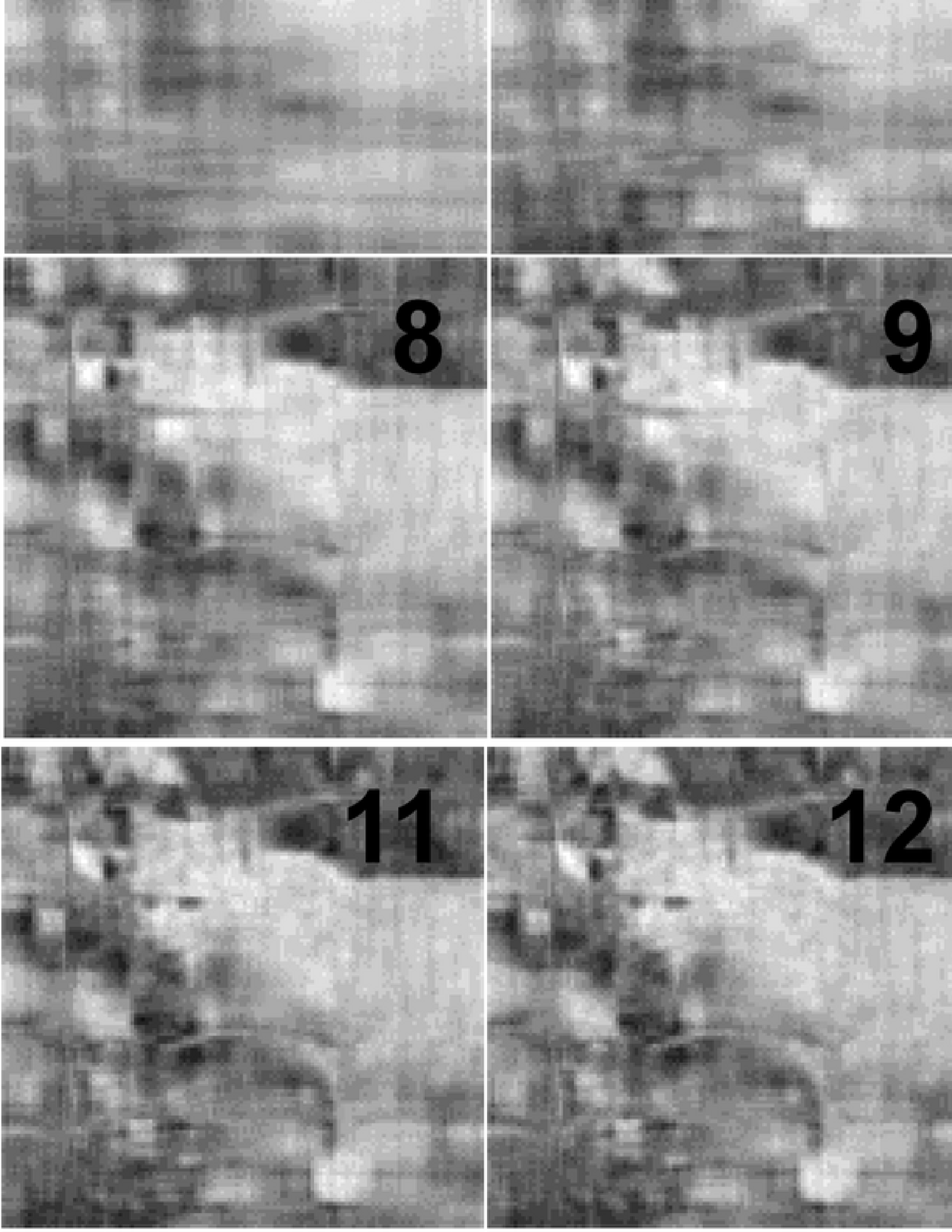}
\end{center}
\caption{Snapshot entropy as a function of $\log\chi$.}
\label{fig04}
\end{figure}

Let us also look at numerical data for different images. We take the image of the Shr\"{o}dinger cat. In this case also, we store $\xi_{i=(x,y)}^{p=1}=\psi_{m=2}(x,y)$ and $\xi_{i=(x,y)}^{p=2}=\psi_{m=L=128}(x,y)$ in the network in advance, and then input $\eta(x,y)=\psi_{\chi}(x,y)$ for $2\le\chi\le L=128$. In this case, separation between two final states of the simulation is perfect: For $\chi\le 7$ the final image is $\xi_{i=(x,y)}^{p=1}$, and for $\chi\ge 8$ the final image is $\xi_{i=(x,y)}^{p=2}$.

\subsection{Ising Snapshot}

\begin{figure}[htbp]
\begin{center}
\includegraphics[width=7.5cm]{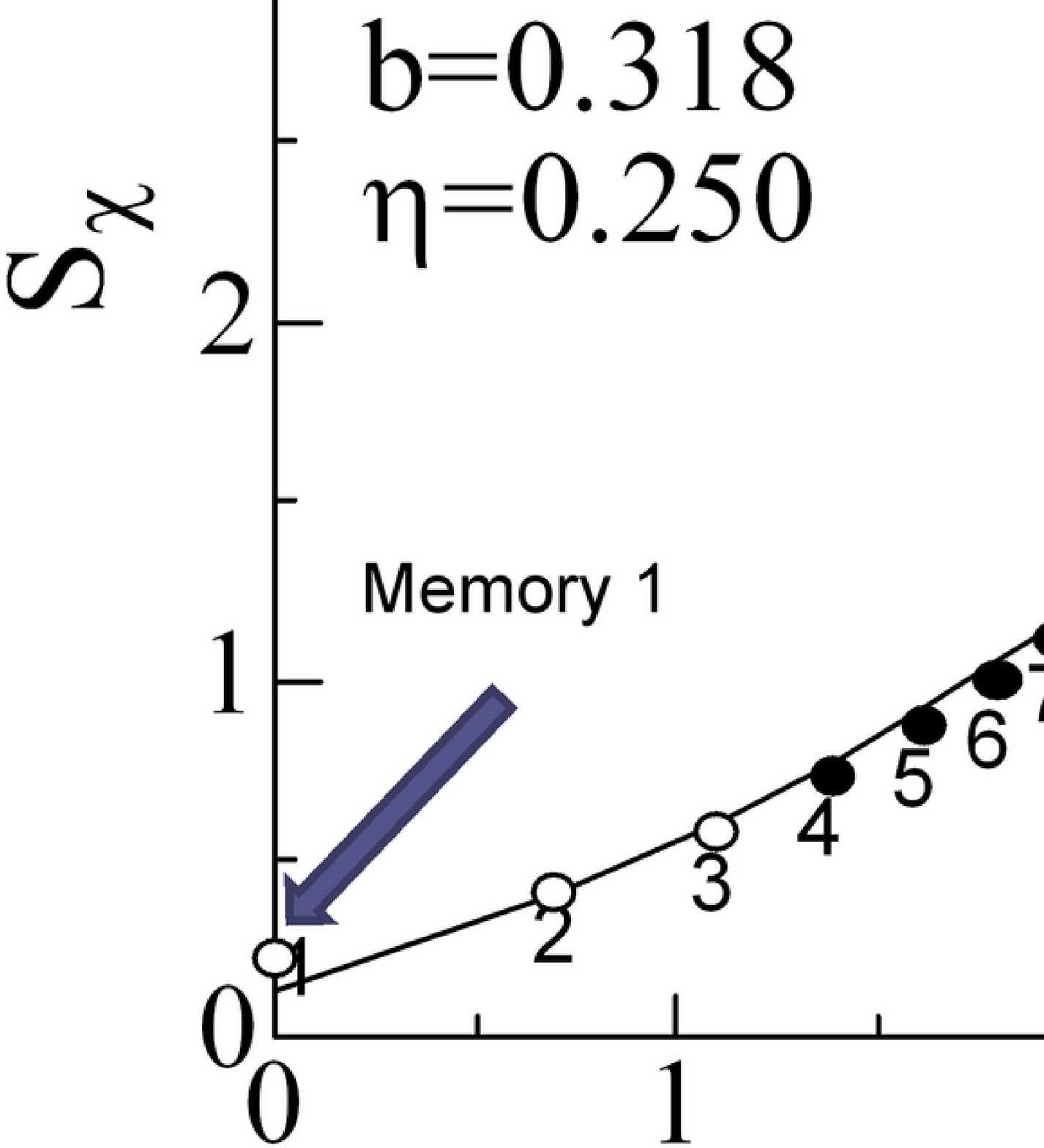}
\includegraphics[width=7.5cm]{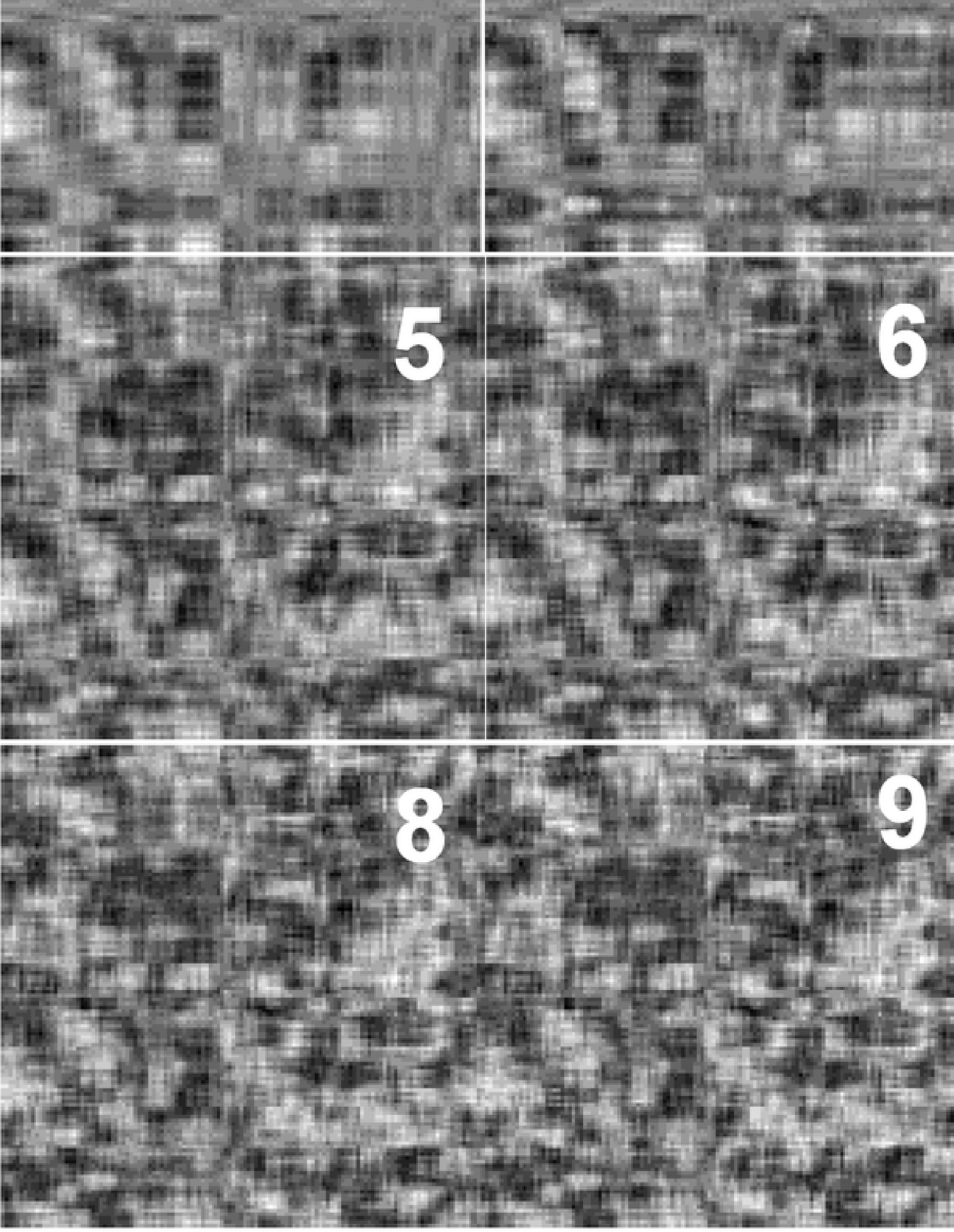}
\end{center}
\caption{Snapshot entropy as a function of $\log\chi$.}
\label{fig05}
\end{figure}

The case of Ising snapshot at $T_{c}$ is also examined. In this case, it has already been found that the entropy scaling is given by
\begin{eqnarray}
S_{\chi}=b\chi^{\eta}\log\left(\frac{\chi}{a}\right),
\end{eqnarray}
with the anomalous dimension $\eta=1/4$ in the scaling theory and constant parameters $a$ and $b$~\cite{Matsueda4,Matsueda6,Imura}.

The numerical data are shown in Fig.~\ref{fig05}. We find that most of all the intermediate data converge to the original image. This is because the Ising snapshot at $T_{c}$ is self-similar, and thus even truncating the image still have some features of scale invariance. This leads to the current result.

\section{Concluding Remarks}

Summarizing, we have evaluated performance of associative memory in a recurrent neural network by based on the SVD method. Our result clearly shows that the ability of the neural network is characterized by the entropy scaling inherent in the SVD: The network can retrieve the original image for the entropy value of the input image beyond the critical value, while the network cannot retrieve it for the input that has the entropy on the logarithmic scaling. In other words, we can introduce the absolute measure of the performance in the neural network by using the SVD.

Now our setup is based on the mutual-coupling-type network, while the network performance is strongly influenced by the RG concept indirectly introduced through the SVD data. In that sense, the present network behaves similarly to hierarchical one like the deep learning machine. The present approach is a hint to unify these different network theories.

Away from the neural network theory, a similar problem has also attracted much attention quite recently in variational methods in condensed matter physics~\cite{Evenbly}. Therein, the tensor-network renormalization state and the multiscale entanglement renormalization ansatz are two important variational wavefunctions at criticality~\cite{Verstraete,Vidal}. The former and the latter are respectively the mutual-coupling type and the hierarchical type, and they are convertible. Their structural analysis may facilitate deeper understanding relationship between the present result and the hierarchical neural networks.

This work was supported by JSPS Kakenhi Grant No.15K05222.

\end{document}